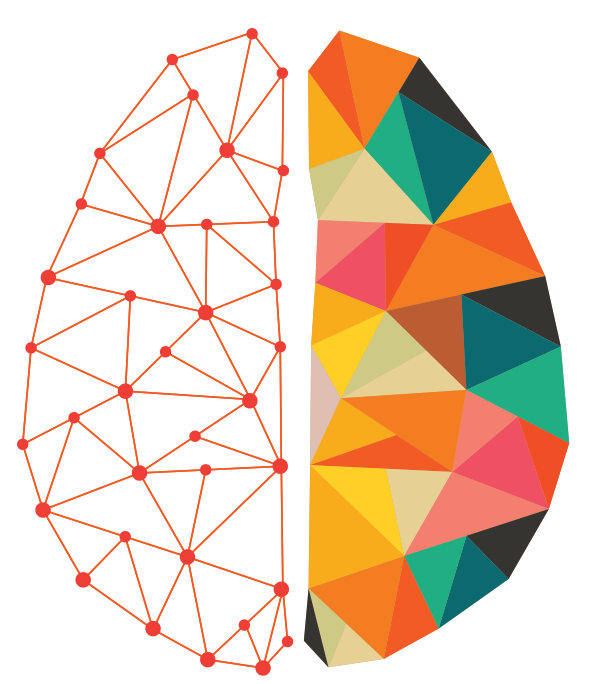

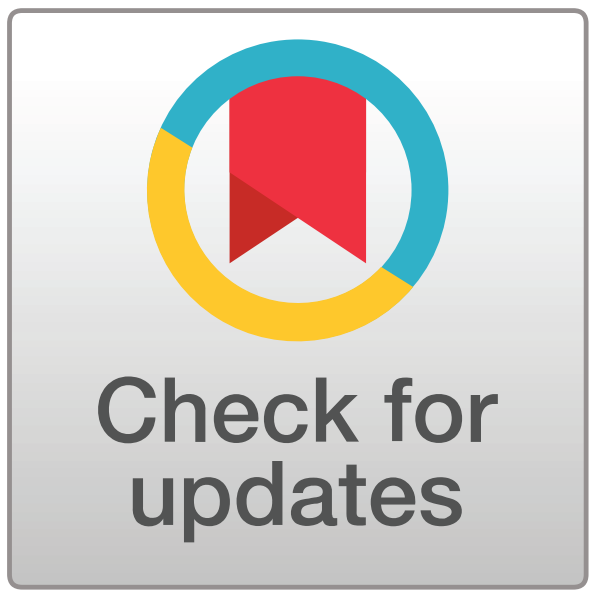

Citation: Roy, O., Moshfeghi, Y., Ibanez, A., Lopera, F., Parra, M. A., & Smith, K. M. (2024). FAST functional connectivity implicates P300 connectivity in working memory deficits in Alzheimer's disease. *Network Neuroscience*, *8*(4), 1467–1490. https://doi.org/10.1162/netn_a_00411

DOI:
https://doi.org/10.1162/netn_a_00411

Received: 22 March 2024
Accepted: 8 August 2024

Competing Interests: The authors have declared that no competing interests exist.

Corresponding Author:
Om Roy
o.roy.2022@uni.strath.ac.uk

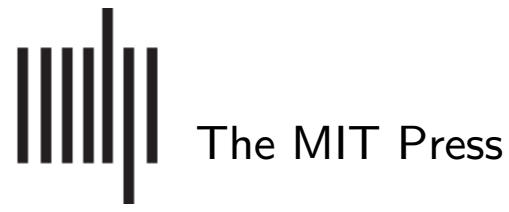

RESEARCH

# FAST functional connectivity implicates P300 connectivity in working memory deficits in Alzheimer's disease

**Om Roy[1], Yashar Moshfeghi[1], Agustin Ibanez[2,3], Francisco Lopera[4], Mario A. Parra[5], and Keith M. Smith[1]**

[1]Computer and Information Sciences, University of Strathclyde, Glasgow, UK
[2]Latin American Brain Health Institute (BrainLat), Universidad Adolfo Ibañez, Santiago, Chile
[3]Global Brain Health Institute, Trinity College Dublin, Dublin, Ireland
[4]Neuroscience Group of Antioquia, Medicine School, University of Antioquia, Medellín, Colombia
[5]Psychological Sciences and Health, University of Strathclyde, Glasgow, UK

**Keywords:** Alzheimer's disease, P300, Working memory, Dynamic functional connectivity, EEG, FAST

## ABSTRACT

Measuring transient functional connectivity is an important challenge in electroencephalogram (EEG) research. Here, the rich potential for insightful, discriminative information of brain activity offered by high-temporal resolution is confounded by the inherent noise of the medium and the spurious nature of correlations computed over short temporal windows. We propose a methodology to overcome these problems called filter average short-term (FAST) functional connectivity. First, a long-term, stable, functional connectivity is averaged across an entire study cohort for a given pair of visual short-term memory (VSTM) tasks. The resulting average connectivity matrix, containing information on the strongest general connections for the tasks, is used as a filter to analyze the transient high-temporal resolution functional connectivity of individual subjects. In simulations, we show that this method accurately discriminates differences in noisy event-related potentials (ERPs) between two conditions where standard connectivity and other comparable methods fail. We then apply this to analyze an activity related to visual short-term memory binding deficits in two cohorts of familial and sporadic Alzheimer's disease (AD)-related mild cognitive impairment (MCI). Reproducible significant differences were found in the binding task with no significant difference in the shape task in the P300 ERP range. This allows new sensitive measurements of transient functional connectivity, which can be implemented to obtain results of clinical significance.

## AUTHOR SUMMARY

Filter average short-term (FAST) connectivity is an EEG analysis method that enhances detection of dynamic functional connectivity changes during cognitive events like event-related potentials, effectively handling EEG noise and maximizing temporal resolution. It reduces the required trial numbers for reliable analysis, particularly beneficial for studying tasks such as working memory. FAST connectivity complements traditional methods by focusing on temporal connectivity patterns, showing superior performance in simulations compared with standard measures. Applied to Alzheimer's datasets, it identifies significant differences in brain activity during visual short-term memory tasks, highlighting its potential for understanding neurological conditions.

## INTRODUCTION

Network science approaches to the analysis of complex networks provide useful tools for the analysis of connectivity between agents (Smith et al., 2017, 2019). The brain is an example of a complex network where pairwise dependencies between brain regions are of value in the detection of cognitive phenomena. It is found that it is neither the spatial nor temporal localization of brain activity that underpins cognitive phenomena and the corresponding brain function but, in fact, how the different areas of the brain are dynamically interconnected over time (Britz et al., 2010; Sporns et al., 2005; Zhao et al., 2022). This has led to a boom in studies of functional connectivity of brain activity across viable formats (Brookes et al., 2011; Bullmore & Sporns, 2009; Stam, 2014)—mainly the BOLD signal in fMRI (Sanz-Arigita et al., 2010) and electromagnetic recordings from EEG and Magnetoencephalogram (MEG). Here, typically, signals from parcellated regions (in fMRI) or sensors (in EEG/MEG) are subject to pairwise measures of connectivity, such as correlation coefficients, coherence measures, or phase-based measures (Canolty & Knight, 2010; Fries, 2015). In particular, there has been a clear increase in the study of functional connectivity changes related to Alzheimer's disease (AD) in the form of the study of the brain as a network using various graph-theoretic tools (de Haan et al., 2009; Stam et al., 2007; Supekar et al., 2008).

The EEG contains important discriminating information relating to sequential brain processes in response to various cognitive tasks (John et al., 1988; Leuchter et al., 1994; Thatcher et al., 2005). Providing a very high-temporal resolution, scalp EEG allows for the direct recording of electromagnetic activity of the brain in a noninvasive, relatively cheap way (Light et al., 2010). Scalp EEG presents several notable limitations, however, with the most prominent being the substantial noise levels inherent in the recorded signals. This noise poses a significant challenge, especially when attempting to investigate the functional connectivity associated with transient cognitive processes occurring within brief time frames, typically spanning mere tens of milliseconds. A pivotal issue within the realm of functional connectivity of EEG signals pertains to the extraction of dependable connectivity estimates within these remarkably short time intervals (Clark et al., 2022). This problem underscores the necessity for novel methodologies to overcome noise-related hurdles and facilitate the precise examination of cognitive processes unfolding at rapid temporal scales (Smith et al., 2019). Measuring dynamic functional brain connectivity in short time windows is gaining increasing recognition in AD research due to its potential to provide information for the early detection of the devastating disease (Johnson et al., 2012; Niu et al., 2019; Paitel et al., 2021). An important reason for this is the growing recognition that intricate changes in brain connectivity can occur before the onset of clinical symptoms; this makes it a promising avenue for early biomarker development and a better understanding of disease progression. AD is not a static condition but involves dynamic changes in brain function; short-time based analysis with noninvasive brain imaging techniques can provide important breakthroughs in AD early detection, especially in low-income countries (Pietto et al., 2016).

Despite the growing popularity of these studies, there has been limited methodological work on the analysis of EEG dynamic functional connectivity (DFC). Previous work typically focuses on the sliding window method (Mateo & Talavera, 2018; Savva et al., 2019; Šverko et al., 2022); while this is fairly effective, the temporal resolution and susceptibility to noise is largely determined by the window size. It has become a priority to simultaneously improve the temporal resolution of DFC while being robust to spurious connections and noise. Methods such as the short-term Fourier transform (Mateo & Talavera, 2018) and wavelet analysis (Li & Chen, 2014) have been frequently applied in this domain, but once again, the dependency on window size causes bottlenecks in regard to temporal resolution and noise robustness.

Dynamic functional connectivity (DFC):
The study of how functional connections in the brain change over short periods, often using methods like sliding windows to capture these variations.

Graph signal processing (GSP) approaches have been employed frequently in the past to perform spectral analysis of signals in the graph domain (Ortega et al., 2018). This is achieved by computing the eigenvalue decomposition of a relevant graph-shift operator such as the graph Laplacian or adjacency matrix followed by the graph Fourier transform. However, the frequencies that emerge through the graph eigenvectors are still determined completely by the graph topology and do not involve the signal itself (Smith et al., 2019).

Graph signal processing (GSP):
A framework for analyzing signals that reside on the vertices of a graph, leveraging the graph structure to perform a spectral analysis.

Here, we propose a new method for extracting reliable estimates of short-term functional connectivity. This is based on a graph-variate signal analysis (Smith et al., 2019), a more general framework for graph signals. Specifically, it describes how to leverage graphs of long-term reliable connectivity information to filter instantaneous bivariate node functions of multivariate signals. In essence, this emphasizes important connections and minimizes spurious ones (a well-known issue in EEG signals). This gives us a readily interpretable method to analyze the transient changes in brain activity at a high-temporal resolution using pairwise connectivity measures between EEG electrodes.

Graph-variate signal analysis:
A general framework for the analysis of multivariate time series on temporally evolving graphs against a stable support.

Graph-variate dynamic (GVD) connectivity (Smith et al., 2019) is when the long-term connectivity estimate that is computed from the signal itself over the given epoch of interest so that the graph signal is directly related to the underlying graph and measurements and, therefore, solely relates to one connectivity function.

Here, we develop and employ a novel methodology based on GVD connectivity that we call filter average short-term (FAST) connectivity. Essentially, FAST connectivity uses the average long-term connectivity matrix over the whole study cohort as a filter of transient functional connectivity at the individual level. Essentially, we are deriving the most consistent connections across all participants and then asking if the temporal activity associated with those connections shows differences between, for example, patients and control.

FAST connectivity:
A method using a global filter computed over all participants as a stable support for the analysis of individual instantaneous EEG connectivity profiles.

Traditional functional connectivity methods typically employ measures such as the amplitude envelope correlation (AEC) or phase locking value (PLV) combined with source reconstruction methods to assess pairwise functional coupling (Hatlestad-Hall et al., 2023). However, the inherent noise in EEG recordings undermines the efficacy of using Pearson correlation coefficients between channel time series as a reliable measure. The FAST filter, which provides a noise-robust matrix representing consistent long-term correlations as a stable support for instantaneous connectivity rather than being the conclusive object of analysis, offers a more appropriate use of this metric for analyzing temporally evolving instantaneous connectivity. This enhances the reliability of coupling measures in the presence of EEG noise.

Pearson correlation coefficient:
A measure of the linear correlation between two variables, used here to calculate long-term connectivity.

EEG recordings are further complicated by individual variability, heterogeneous artifacts, volume conduction effects, and low spatial resolution, which pose significant challenges for spatial filtering approaches such as source reconstruction methods. These methods struggle to accurately solve the inverse problem of mapping a scalp-recorded activity to specific brain regions, especially given the unknown number of sources at any given time. FAST connectivity does not operate in this more traditional domain of analysis and is not intended to replace spatial methods for identifying precise brain regions of activity. Instead, it generalizes consistent statistical dependencies across broader brain regions and serves as a *temporal* filtering technique. This makes it a valuable complement to spatial filtering approaches, enhancing the overall analysis of brain connectivity.

As we shall see, the high-temporal resolution of brain activity provided by the EEG (Light et al., 2010) can now be exploited to detect more sensitive and specific cognitive changes in very short time frames.

Event-related potentials (ERPs):
Brain responses that are the direct result of a specific sensory, cognitive, or motor event.

We demonstrate the power of FAST connectivity in simulations for picking out the true activity of event-related potentials (ERPs) in the presence of different levels of noise and different numbers of trials. We then apply this to the dataset containing EEG signals from the participants in the visual short-term memory (VSTM) tasks (Smith et al., 2022). Following this, we perform rigorous statistical testing on temporal windows resulting from the multilayer graph-variate tensor. This uncovers a potential biomarker for the early detection of AD.

## METHODS

### *Background*

Modular Dirichlet energy (MDE):
A measure previously used to analyze brain connectivity, similar to the approach in FAST connectivity.

The method proposed is inspired from the modular Dirichlet energy (Smith et al., 2017) and the graph-variate signal analysis (Smith et al., 2019) methods. We thus briefly introduce these concepts.

The Dirichlet energy of a graph signal $\mathbf{x}$ is defined as follows:

$$\mathbf{x}^T L \mathbf{x} = \sum_{i,j=1}^{n} w_{ij} \left( x_i - x_j \right)^2 \tag{1}$$

(Smith et al., 2017).

Essentially, this allows us to contrast pairwise graph signal smoothness or variability with a measure $w_{ij}$.

Local Dirichlet energy:
A measure used in FAST connectivity to analyze the variability within localized regions of brain activity.

The squared difference between signal pair values can be considered as a localized measure of the variation between signal pair values. This captures the local variation of the signal. A higher value would indicate higher variation in the signal pair region, whereas if it was small, the signal pair values are fairly constant or change smoothly in the localized region. The Dirichlet energy captures the sum of the local variations over the graph. The term *local* Dirichlet energy will be used from now to refer to pairwise or modular squared differences.

Graph-variate signal analysis is defined formally as follows:

$$\left( \mathbf{W} \circ \underline{\mathbf{J}}_{(t)} \right)_{ij} = \begin{cases} w_{ij} F_V \left( x_i(t), x_j(t) \right), & \text{if } i \neq j \\ 0, & \text{if } i = j \end{cases} \tag{2}$$

where the formula defines the bivariate analysis of the multivariate signal $X$ filtered by the corresponding static matrix $W$ of the graph-variate signal. $\underline{\mathbf{J}}_{(t)}$ denotes the $t$'th $n \times n$ matrix of $\underline{\mathbf{J}}$, and $\circ$ is the Hadamard product. Each timestep of $\underline{\mathbf{J}}$ is defined by a $n \times n$ matrix computed using the pairwise bivariate connectivity values between signal pairs. The form of dynamic connectivity is determined by the node function $F_V$.

GVD connectivity is defined as a graph-variate signal analysis in which **W** = **C** is a static adjacency matrix constructed from the long-term stable dependencies of the multivariate signal itself. We define our tensor for analysis from Smith et al. (2019) as follows:

$$\theta(\mathbf{x}_i, \mathbf{x}_j, t) = \begin{cases} c_{ij} F_V(x_i(t), x_j(t)), & \text{if } i \neq j \\ 0, & \text{if } i = j \end{cases} \tag{3}$$

The multilayer network $\theta$ is constructed using different relevant combinations of node functions and long-term stable connectivity pairs.

Each $c_{ij}$ used to construct **C** is constructed using relevant connectivity measures that give a reliable estimate for long-term term connectivity. A standard approach is the Pearson correlation coefficient computed over the whole epoch of interest:

$$c_{ij} = \frac{\sum_{t \in T}(x_i(t) - \bar{\mathbf{x}}_i)(x_j(t) - \bar{\mathbf{x}}_j)}{\sqrt{\sum_{t \in T}(x_i(t) - \bar{\mathbf{x}}_i)^2}\sqrt{\sum_{t \in T}(x_j(t) - \bar{\mathbf{x}}_j)^2}}, \tag{4}$$

(Smith et al., 2019), where $T$ is the epoch of interest and $\bar{\mathbf{x}}_i$ is the mean of the values over time of the node $i$ and where $T$ is the epoch of interest and $\bar{x}_i$ is the mean of the values over time of the node $i$; combining this with the squared difference, GVD connectivity can be defined as follows:

$$\theta(\mathbf{x}_i, \mathbf{x}_j, t) = c_{ij}(\tilde{x}_i(t) - \tilde{x}_j(t))^2, \tag{5}$$

where $\tilde{x}_i(t)$ is the normalized signal over the node space:

$$\tilde{x}_i(t) = \frac{x_i(t) - \bar{\mathbf{x}}(t)}{\sqrt{\frac{1}{n-1}\sum_{k=1}^{n}(x_k(t) - \bar{\mathbf{x}}(t))^2}}, \tag{6}$$

and $\bar{\mathbf{x}}(t)$ is the mean over nodes of the signal at time $t$:

$$\bar{\mathbf{x}}(t) = \frac{1}{n}\sum_{k=1}^{n} x_k(t). \tag{7}$$

It is clear now that the reformulated Dirichlet energy is a special case of a graph-variate signal analysis.

#### *FAST Connectivity*

Time-locked cognitive task:
An experimental design where participants perform tasks that are synchronized to specific time points to measure brain activity.

We now present FAST connectivity. A single filter is proposed for all participants in time-locked cognitive task-based experiments. The filter takes the long-term connectivity estimates of all participants in the experiments and averages over them to create a single FAST filter for all participants that automatically emphasizes important connections and suppresses spurious ones in the general time-locked cognitive task of interest (in this case, the VSTM binding and shape tasks). We define the FAST filter as follows:

**Definition 1. FAST Filter**: *Where C is the matrix of the absolute values of the individual long-term correlation estimates, with $c_{ij}$ representing each entry in the matrix. For P = 1, 2 … N, where P is each participant and N is the total number of participants. We define our FAST filter as follows:*

$$c_{ij}^{FAST} = \frac{\sum_{P=1}^{N} c_{ij}^{P}}{N} \tag{8}$$

$$c_{ij} = \left| \frac{\sum_{t \in T} (x_i(t) - \bar{\mathbf{x}}_i)(x_j(t) - \bar{\mathbf{x}}_j)}{\sqrt{\sum_{t \in T} (x_i(t) - \bar{\mathbf{x}}_i)^2} \sqrt{\sum_{t \in T} (x_j(t) - \bar{\mathbf{x}}_j)^2}} \right|, \tag{9}$$

We have defined our long-term connectivity estimate as the modulus of the Pearson correlation coefficient; this captures the long-term stable magnitude of the correlation of all participants in the task. Following Definition 1, we define FAST connectivity as follows:

**Definition 2. FAST Connectivity:** *For each P = 1 … N where N is the total number of participants, the same FAST filter is applied to each participant. Fast connectivity is the analysis of the network tensor of the form.*

$$\theta^{\text{FAST}}\left(\mathbf{x}_i^P, \mathbf{x}_j^P, t\right) = \begin{cases} c_{ij}^{\text{FAST}}\left(\tilde{x}_i^P(t) - \tilde{x}_j^P(t)\right)^2, & \text{if } i \neq j \\ 0, & \text{if } i = j \end{cases} \tag{10}$$

FAST connectivity proposes the same filter for all participants *P*. This filter is constructed using the magnitude of the stable long-term correlation averaged over all participants.

Setting $w_{ij}$ as the relevant entry of the FAST filter matrix allows us to weigh corresponding instantaneous variability by long-term stable connections. In layman's terms, we first determine which brain regions are most consistently strongly connected in terms of statistical dependencies over the whole cohort, we are then "focusing" on these regions and analyzing the variability in these localized regions using the squared difference function.

Overall, the FAST connectivity analysis is sensitive to both fine-scale variations within individual EEG signals and broader patterns shared across participants. By combining a global measure with local information, the method is effective in identifying regions that not only vary locally but also exhibit strong synchronized connectivity across participants. This should reflect meaningful functional connectivity patterns while reducing noise and spurious connectivity.

Using the absolute value of the long-term correlation coefficient for the global filter avoids cancelling out information from important connections in network averages.

Similar to the modular Dirichlet energy (MDE; Smith et al., 2017), a prototype of GVD connectivity, FAST connectivity analyzes the temporal brain networks from a unique angle compared with other approaches such as time series analysis of network metrics. Essentially, one stable network of long-term connectivity is computed over the whole epoch and averaged over all participants; this is used as a support for localized analysis of very small temporal windows, allowing us to maximize the high-temporal resolution of EEG signals. The activity is, in fact, encoded in the graph signal itself on the temporally evolving edge weights. This allows for the analysis of smaller temporal windows of activity and also the analysis of the overall long-term activity.

Temporal brain networks: Networks representing brain connectivity that evolve over time, analyzed in FAST connectivity.

However, one clear limitation of this approach is that we lose out on potentially important short-term connectivity between otherwise unimportant long-term connections. We do not assume that such information is not important, but rather the problem of spurious correlations over short-temporal windows far overshadows it. The success of the method demonstrated in simulations and real data backs this argument.

#### Network Analysis of FAST Connectivity

FAST connectivity can be computed over arbitrarily selected windows, resulting in a connectivity matrix for each window. We can then straightforwardly compute a network analysis on these connectivity matrices. Here, we use the mean edge weights as well as an average local weighted clustering coefficient of FAST functional connectivity. The mean edge weight is computed as follows:

$$\bar{W}(t) = \frac{1}{n^2}\sum_{i=1}^{n}\sum_{j=1}^{n}\Delta_{ij}. \tag{11}$$

We computed the average local weighted clustering coefficient for each temporal window as follows:

$$C_{avg}(t) = \frac{1}{n}\sum_{i=1}^{n}\sum_{j,k=1}^{n}\Delta_{ijt}\Delta_{ikt}\Delta_{jkt} = \frac{1}{n}\sum_{i=1}^{n}\left(\underline{\Delta}^{3}_{(t)}\right)_{ii} \tag{12}$$

(Smith et al., 2019).

We limited our choice to these two network metrics as we did not perform any binarization on the connectivity profiles in order to maximize the information we can gain. Thus, we are just analyzing changing edge weights of the completely connected graph; as a consequence of this, the weighted clustering coefficient and the edge weights represented the general topological distribution of the connectivity profile effectively and provided sufficient results.

Furthermore, at this initial stage of our methodological development, our goal is to identify significant changes in global functional connectivity. To achieve this, we are not conducting a node-by-node analysis. This approach is crucial for small sample-sized datasets, as nodal-level testing greatly reduces statistical power due to multiple comparison corrections. Instead, we are adopting a hypothesis-free approach regarding specific brain regions of interest.

The mean of the edge weights in this scenario essentially captures the Dirichlet energy of the entire instantaneous filtered connectivity profile that provides a reliable measure of instantaneous connectivity as indicated in previous studies (Smith et al., 2017).

The clustering coefficient, on the other hand, quantifies the number of connected triangles in a network and, thus, the tendency of nodes to cluster together. This weighted version multiplies the triangle weights together, with larger values where all triangle weights are large. The average value for each temporal window emphasizes the strongly clustered components in the signal. The computation is fairly straightforward with the sum of the main diagonal of the cube of the tensor divided by the number of nodes (EEG electrodes).

We can make two interesting observations here, taking into account the law of large numbers stating that as the number of independent samples increases, the empirical mean converges to the true mean; we can conclude that increasing the number of electrodes and, thus, nodes (and thus independent samples) will give us a more stable and reliable estimate of the mean for a given time step, allowing for a greater temporal resolution. This has been implicated in previous studies, where there is strong evidence showing that as a result of reducing electrode density, networks tended to get skewed (Hatlestad-Hall et al., 2023); this effect was most prominent below 64 electrodes. Also, the central limit theorem tells us that the estimate will converge to a normal distribution as the number of independent samples increases, suggesting that increasing the number of electrodes would allow the network metric estimates to follow a more Gaussian distribution, allowing us to exploit the various statistical approaches that assume Gaussianity.

#### *Wavelet Power Spectra Analysis*

The wavelet transform (Kumar et al., 2022) is a powerful tool for analyzing temporally varying signal data. It allows for the decomposition of a signal into components localized in both time and frequency domains. One of the commonly used wavelets for continuous wavelet transform (CWT) is the Morlet wavelet, which is particularly useful for detecting oscillatory patterns in the signal.

The CWT of a signal $x(t)$ using a mother wavelet $\psi(t)$ is defined as follows:

$$W_x(a,b) = \frac{1}{\sqrt{a}} \int_{-\infty}^{\infty} x(t)\psi^* \left(\frac{t-b}{a}\right) dt, \tag{13}$$

where $a$ is the scale parameter, which controls the dilation of the wavelet; $b$ is the translation parameter, which controls the translation of the wavelet; and $\psi^*(t)$ is the complex conjugate of the mother wavelet $\psi(t)$.

The Morlet wavelet is defined as follows:

$$\psi(t) = \pi^{-\frac{1}{4}} e^{j\omega_0 t} e^{-\frac{t^2}{2}} \tag{14}$$

where $\omega_0$ is the central frequency of the wavelet.

The wavelet transform can be viewed as a convolution of the signal $x(t)$ with a set of wavelet functions. Each wavelet function is a scaled and shifted version of the mother wavelet $\psi(t)$. The wavelet coefficients $W_x(a, b)$ represent the correlation between the signal and the wavelet at different scales $a$ and positions $b$.

To analyze the EEG signals, we split the data into 10 disjoint temporal windows. For each window, we computed the power spectrum for each channel using the wavelet coefficients. This allowed us to analyze the signal in the frequency-time domain. This process helps in understanding how the power of different frequency components of the signal varies over time. In this case, we used the Morlet wavelet as our mother wavelet of choice.

#### *Simulations*

We utilized open-source MATLAB functions provided by Yeung et al. (2004, 2007) to generate the simulated EEG data. The simulated data consist of two key components: The signal component is generated to mimic the power spectrum of a typical human EEG recording. The peak component is parameterized to describe the position of the center of the peak or ERP, its frequency, and its amplitude. These parameters enable us to create sample ERPs, which serve as the basis for testing the effectiveness of our method. The EEG simulation functions provide a setup with 31 electrodes, each sampled at a frequency of 200 Hz, with an epoch duration of 0.8 s. To generate independent samples, we averaged the random signals over varying numbers of trials, resulting in single 31 × 200 dimensional samples that closely resemble real EEG data.

For our experimental setup, we aimed to replicate conditions akin to typical comparisons between participants in time-locked VSTM tasks. We created 20 independent samples consisting solely of EEG time series, aligning with the EEG power spectrum of a typical human. In parallel, we generated 20 independent samples with specific ERPs, including the N100 and P300 components. The amplitude of the general EEG signal was set at 10.

The N100 component was parameterized with an amplitude of −5 (typically negative) and a frequency of 15 Hz, with a center position at 25 frames (around 100 ms) considering the 0.8-s epoch. The P300 component was parameterized with an amplitude of 5 (positive and larger than N100) and a frequency of 5 Hz, with a center position at 75 frames (around 300 ms) within the 0.8-s epoch. To introduce a realistic variability, the functions incorporate temporal jitter at the onset of the ERPs, mirroring the kind of activity observed in actual EEG ERP data. We then added random Gaussian white noise to the samples containing the simulated ERPs to test the robustness of our FAST filtering method in the presence of variable levels of external noise. This approach allows us to rigorously test the performance of our method in distinguishing between the presence and absence of these specific ERP components in simulated EEG signals.

The N100 has been previously implicated in various neurological disorders such as schizophrenia and attention-deficit/hyperactivity disorder (ADHD; Murias et al., 2007). The P300 is characterized by a positive deflection in the EEG signal and usually occurs around 300 ms post the presentation of stimuli. The P300 can be influenced by the given task the participant is involved in and is associated with the evaluation of the relevance of stimuli. The P300 has been heavily researched in AD (Paitel et al., 2021; Parra et al., 2012; Pedroso et al., 2012) and is associated with decision-making and working memory.

#### VSTM Data

This study examines patients with mild cognitive impairment (MCI) due to AD, focusing on the predementia stage. The MCI participants are categorized into familial (MCI-FAM) and sporadic (MCI-SPO) groups. Familial AD participants exhibit AD symptoms but do not yet meet the criteria for dementia, although they will inevitably progress to it. Sporadic AD participants, representing the most common type, have an undetermined risk of developing dementia. Both groups are compared with control participants without genetic mutations and free of psychiatric or neurological disorders.

All participants gave written informed consent following the Helsinki Declaration. The Ethics Committees of the Institute of Cognitive Neurology (INECO) and the University of Antioquia approved the study.

**Sporadic MCI.** Table 1 examines sporadic MCI (MCI-SPO) focusing on demographic and clinical characteristics of the subjects.

We report on the Mini-Mental State Examination (MMSE), with a detailed clinical and neuropsychological profile available in Pietto et al. (2016). Patients exhibited multiple-domain amnestic MCI based on various tests. Nine patients were at high risk for AD conversion, while three had nonamnestic MCI multidomain. The data include a 128-channel EEG activity recorded at 512 Hz using a Biosemi Active Two System, filtered from 1 to 100 Hz, and down-sampled to 256 Hz.

**Table 1.** Demographic and clinical characteristics of subjects MCI-SPO

| | **MCI patients (*n* = 13)** | **Healthy controls (*n* = 19)** |
|---|---|---|
| Age (years) | 73.08 ± 9.01 | 67.21 ± 10.14 |
| Education (years) | 14.08 ± 4.44 | 16.50 ± 1.99 |
| MMSE scores | 26.46 ± 2.47 | 29.50 ± 0.52 |

**Familial AD dataset.** The MCI-FAM group carries the E280A mutation of the presenilin-1 gene, leading to guaranteed early-onset familial AD. Table 2 details the basic demographic and clinical characteristics as with MCI-SPO.

The data consist of a 60-channel EEG activity recorded with a 64-channel EEG cap using SynAmps 2.5 in Neuroscan at 500 Hz, band-pass filtered from 1 to 100 Hz with impedances below 10 k. Four ocular channels were discarded after being used to factor out oculomotor artifacts.

**VSTM binding and shape task description and performance.** In the assessment of VSTM, two distinct tasks are employed (Smith et al., 2022): a shape-only change detection task and a binding task. In the shape-only task, participants are presented with arrays featuring three different black shapes, while in the binding task, the arrays consist of three distinct shapes, each with a unique color. Each trial in both tasks comprises three phases: an initial encoding period (lasting 500 ms), during which participants view a study array on the screen, followed by a short delay of 900 ms, and concluding with the test period. In the test period, a test array is displayed, and participants are tasked with determining whether the objects in the two arrays are identical or different. To prevent reliance on spatial cues, the positions of objects are randomized. Shapes and colors are randomly selected for each trial from sets of eight options. Notably, in 50% of the trials, both arrays feature identical objects. In the remaining 50%, changes occur: In the shape task, two shapes are substituted with new ones, while in the binding task, the colors of two shapes are interchanged. Participants commence with a practice session and subsequently complete 100 trials for each task. Importantly, the order in which they engage in the binding and shape tasks is systematically counterbalanced across participants, ensuring a comprehensive exploration of VSTM dynamics (Pietto et al., 2016; Smith et al., 2022).

The response accuracy to the two VSTM task conditions was similar for both controls (Mann–Whitney $U$: 34, $Z = 1.17$, $p = 0.24$, $d = 0.64$) and MCI-FAD (Familial Alzheimer's Disease) patients (Mann–Whitney $U$: 28, $Z = 1.63$, $p = 0.10$, $d = 0.77$). However, between-group comparisons showed that controls had higher accuracy in the shape–color binding condition (Mann–Whitney $U$: 22.5, $Z = -2.08$, $p < 0.05$, $d = 0.93$), with no significant differences observed in the shape-only condition (Mann–Whitney $U$: 25.0, $Z = -1.89$, $p = 0.063$, $d = 1.02$).

Within-group comparisons showed no significant differences in the response accuracy between task conditions for controls (Mann–Whitney $U$: 66.5, $Z = 1.42$, $p = 0.16$, $d = 0.60$) or MCI-SPO patients (Mann–Whitney $U$: 54, $Z = 1.54$, $p = 0.12$, $d = 0.64$). MCI-SPO patients performed significantly worse than controls in both the shape-only (Mann–Whitney $U$: 42.5, $Z = 2.33$, $p < 0.05$, $d = 0.91$) and shape–color binding (Mann–Whitney $U$: 42.0, $Z = 2.35$, $p < 0.05$, $d = 0.92$) conditions.

**Table 2.** Demographic and clinical characteristics of subjects MCI-FAM

| | **Patients ($n$ = 10)** | **Healthy controls ($n$ = 10)** |
|---|---|---|
| Age (years) | 44.4 ± 3.2 | 44.3 ± 5.6 |
| Education (years) | 7.3 ± 4.1 | 6.8 ± 2.9 |
| MMSE scores | 25.20 ± 4.50 | 29.10 ± 1.10 |

Signal preprocessing was performed to get signals band-passed into delta (0.01–4 Hz), theta (4–8 Hz), alpha (8–12 Hz), beta (12–30 Hz), and gamma (>30 Hz) frequencies with each epoch lasting 1 s poststimuli exposure.

### Statistical Methods

We designate as a result of clinical interest as a given temporal window where there is a significant difference between patients and controls in the binding task and no significant difference in the shape task as this points to a specific binding deficit in AD.

The network analysis was done using FAST connectivity. We first applied Definition 1 to create our FAST filter of overall general VSTM task activity. A separate filter was created for the MCI-FAM dataset and for the MCI-SPO dataset due to them having different experimental parameters (i.e., number of electrodes). We then split the FAST connectivity tensors into ten 0.1-s nonoverlapping temporal windows by averaging over smaller time windows to give us 10 matrices of FAST connectivity for each participant with a high-temporal resolution.

Nonparametric Wilcoxon rank-sum tests are performed to assess for statistical significance between patients and controls. These are computed at each 0.1-s temporal window between the vectors of mean network metrics for patients and controls at each temporal window. This is repeated for the mean edge weights and the mean weighted clustering coefficient values. The direction and size of the differences are calculated using Cohen's $d$ effect size. In our experiments, a negative value would indicate a greater local Dirichlet energy in the AD patients for the given VSTM task.

In order to account for the multiple comparisons, we applied the Benjamini-Hochberg false discovery rate (FDR) correction to account for multiple temporal significance testing. This was done at the 10% and 5% level. While 5% is often held as the strict standard, the 10% level allows us to look for sensitivity pointing toward repeatability across datasets, that is, where one dataset passed FDR at a given time point at 5% and the other at 10%.

## RESULTS

### FAST Connectivity Outperforms Wavelet Transform and Unfiltered Connectivity at Robust, Temporally Precise ERP Detection

A simulation in a general case scenario is implemented to show the effectiveness of our method at picking up relevant ERPs. This is tested across varying noise levels and the number of task trials. First, we performed a three-way comparison between the unfiltered node function, individual GVD filters, and our FAST filter. Our aim is to evaluate the effectiveness of our methodology in detecting these simulated ERPs compared with a "control" group where the ERPs are not present.

Gaussian white noise is added randomly to each of the electrodes equally in the simulated setup (note, we are adding a random white noise, which is distinct from the signal generated by the MATLAB functions that resemble the power spectrum of a typical human EEG recording).

Figure 1 shows visually the effect of the FAST filter on instantaneous EEG connectivity profiles. We used a time step consisting of a simulated ERP in the presence of a high level of

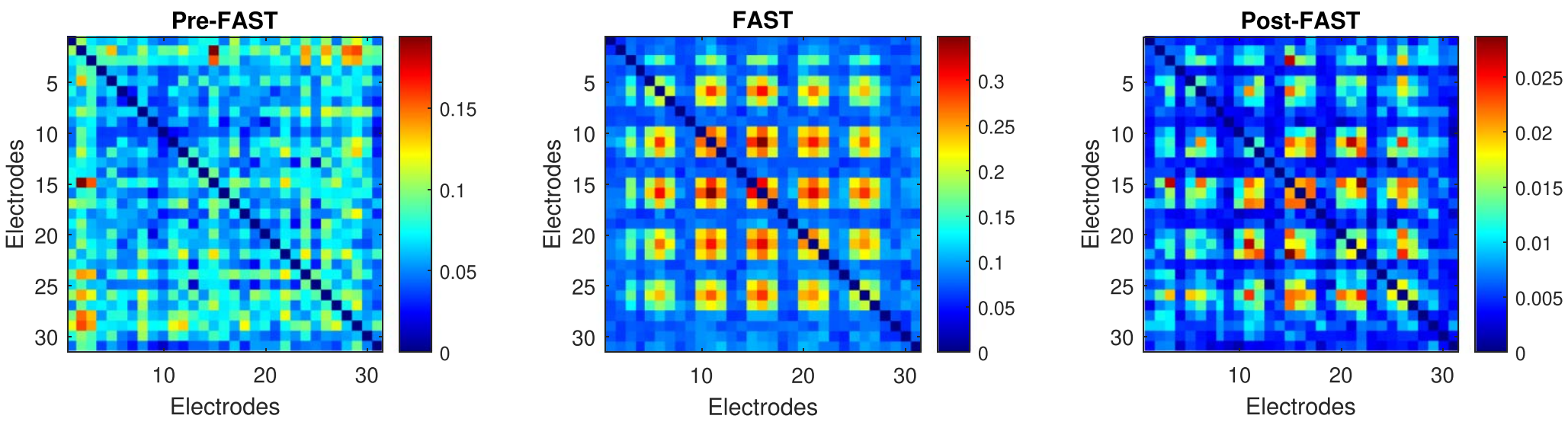

**Figure 1.** A visual demonstration of the effect of the FAST filter on the EEG instantaneous connectivity profile. The left panel shows the raw EEG connectivity profile, while the right panel displays the connectivity profile after applying the FAST filter, illustrating the filter's ability to enhance relevant connectivity patterns and reduce noise.

Gaussian distributed noise. We can see that the FAST filter has first established the highly important connections in terms of long-term stable connectivity and identified regions of lower consistency where spurious connections are likely to be present. After the application of the FAST filter, we can see that areas of importance have been emphasized (red patches) while spurious connections and noise have been weighted down. Note that the FAST filter is *task selective* and that only areas that are both instantaneously and globally consistent are emphasized; that is, if we had a very weak connection instantaneously but it was strong globally, it would not be emphasized. This is a justification for us using both the shape and binding tasks to compute our FAST filter as the FAST filter would activate binding-specific areas and shape-specific areas dependent on the task being analyzed, while common strongly connected areas in both tasks would be automatically emphasized. This allows us to utilize the stable connectivity information of both tasks while making sure we do not obtain spurious results due to differences in the tasks. While we can do this due to the similarity of the VSTM binding and shape task and given that they are on the same timescale, highly differing tasks on different timescales would likely not be suitable to construct a FAST filter on as the likelihood of spurious results would increase.

Although Graph-Variate Signal Analysis (GVSA) is usually performed with distinct individual filters of long-term stable connectivity distinct for each participant, this would result in individual variation and noise in the EEG signals to bias the instantaneous connectivity profiles resulting in false positives or Type 1 errors.

Figure 2 illustrates this; when we use individual filters, the noise in the medium is resulting in a very high number of Type 1 errors, with a spurious significant difference being picked out (all time steps are significant, with the mean clustering coefficient and edge weights overlapping). FAST connectivity, as shown above, results in no false positives.

Common baseline EEG analyses revolve around the analysis of the power spectrum of the frequency domain of EEG signals. Therefore, it is important to study the added benefits of our approach for classification of EEG signals beyond the power spectral analysis. In this vein, we compare FAST functional connectivity with the wavelet transform, a common approach to analyzing temporally varying signal power spectra. The CWT produces wavelet coefficients, which capture how the signal correlates with the wavelet at different scales (frequencies) and translations (times). The wavelet coefficients obtained from the CWT are complex numbers. The magnitude squared of these coefficients represents the power of the signal at different frequencies and times.

We compute the overall power of the signal within each temporal window across all channels. We then computed this for all simulated participants or samples with and without a simulated ERP and performed the rank-sum test for significance with FDR correction at varying levels of trials and external noise levels.

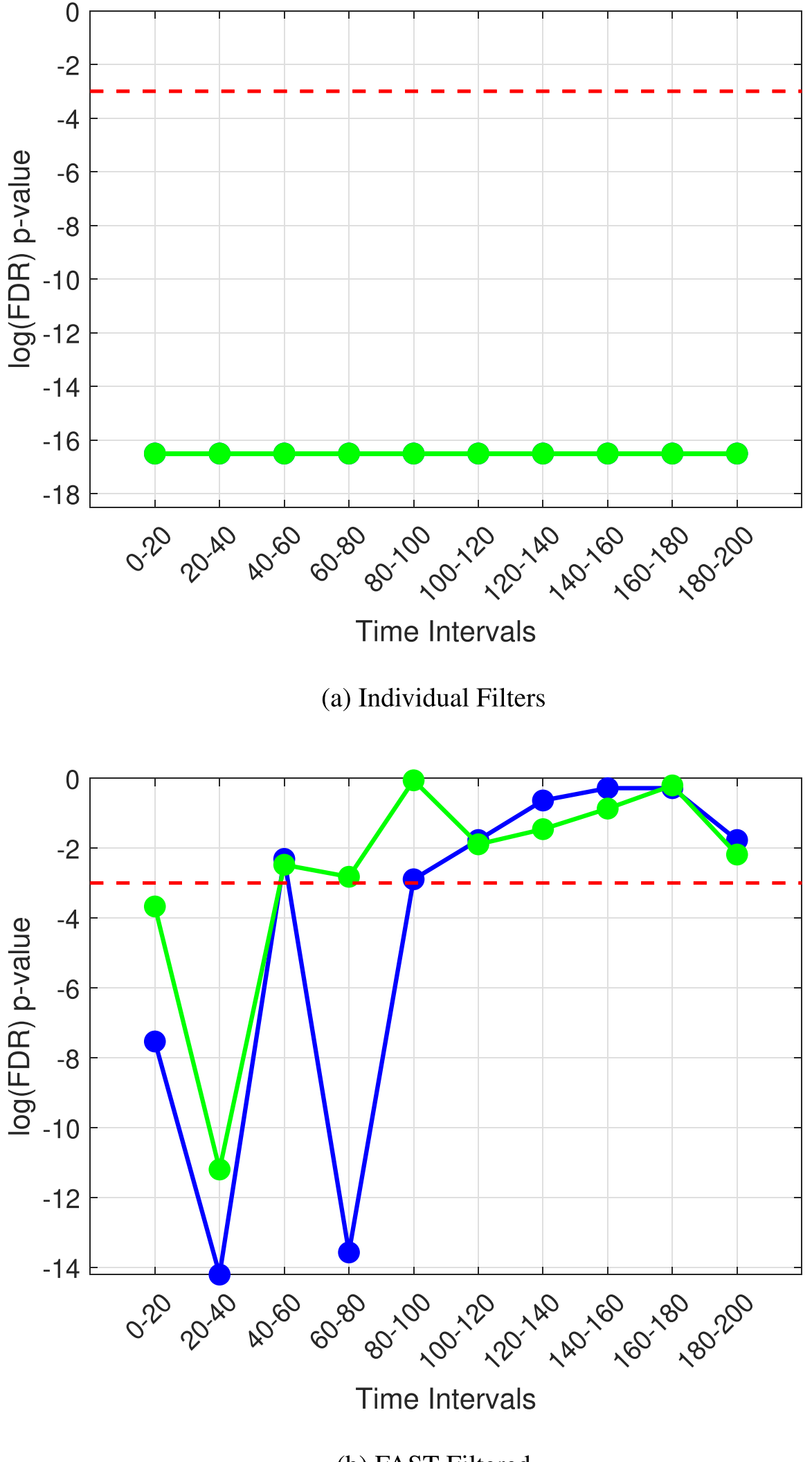

(a) Individual Filters

(b) FAST Filtered

**Figure 2.** The *p* value plots over time for the detection of the N100 and P300 ERP's (individual filters (A) vs. FAST filter (B)).

We decided to test FAST connectivity's robustness to noise more rigorously by varying levels of added Gaussian noise at different numbers of trials for each simulated participant EEG. We extracted the FDR-corrected *p* values at the P300 ERP time steps of interest (predetermined to exist at these time steps), enabling us to compare the ability of unfiltered dynamic connectivity (the squared difference node function with a support of 1s with 0s on the main diagonal as the "filter") against the FAST filtered approach to pick out significant differences at these points while also making a comparison with the more traditional wavelet power spectra approach. We tried trials ranging from 50 to 300 corresponding to typical real-life experiments where EEG signals are recorded.

Figures 3A, 3B, and 3D show the *p* values obtained using unfiltered functional connectivity, FAST functional connectivity, and wavelet analysis, respectively. Our FAST functional connectivity approach consistently and strongly outperforms the baseline methods in almost all cases.

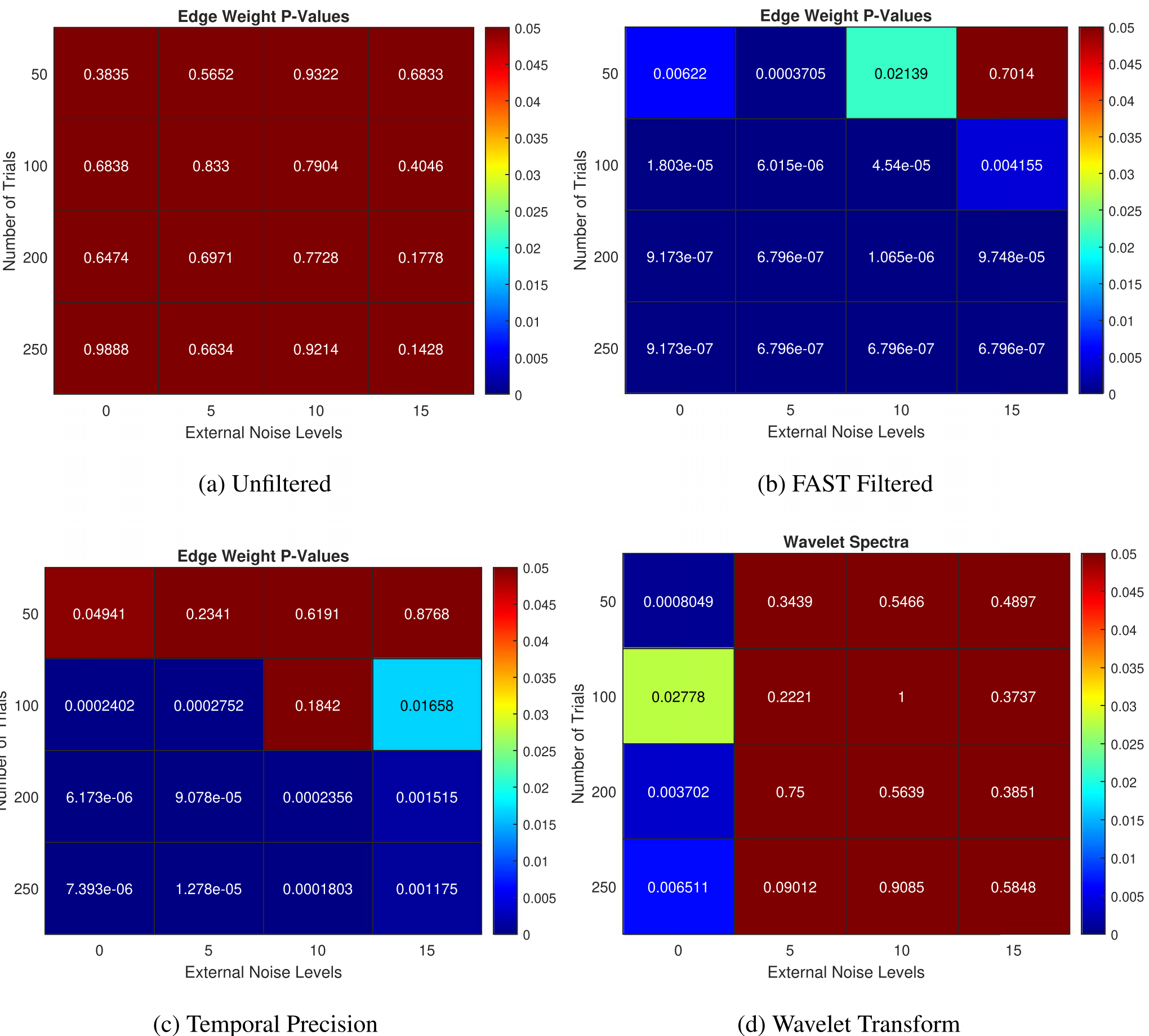

(a) Unfiltered

(b) FAST Filtered

(c) Temporal Precision

(d) Wavelet Transform

**Figure 3.** The *p* values at a predetermined time step of the simulated P300 at increasing levels of trial size and external Gaussian noise. (A) Unfiltered, (B) FAST filtered, (C) temporal precision after applying the FAST filter, and (D) wavelet spectral analysis.

The wavelet power spectra analysis, at first glance, performs adequately, being able to detect the ERP at all trial sizes when no external noise is added. On further inspection looking at the trend of *p* values, the lowest *p* value comes at 50 trials; this is an unexpected behavior, and the expected pattern emerges at 100 trials and above. This is a strong indication of a Type 1 error taking place due to the inherent noise in the EEG signal causing spurious "significant differences." Upon the further observation of the *p* values of the wavelet transform at all time steps, we noticed that while the ERP could be detected at an external noise level of 0, there was a large number of false positives at other non-ERP time steps, reducing the power of our statistical analysis greatly. Referring back to Figure 2, this is a similar behavior when the individual filters of GVSA are used. Importantly, we noticed that FAST connectivity had a very low (almost 0) false positive rate, with significant differences only at the time steps of interest.

Another important consideration in analyzing DFC in small temporal windows is the temporal resolution we can achieve while still maintaining robustness to noise. Previous methods depended heavily on the length of the sliding window in finding a trade-off between temporal resolution and robustness to noise (Savva et al., 2019). We decided to test the window length dependency of FAST connectivity by setting the number of windows equal to the sampling rate, that is, maximum temporal resolution and repeating our varying trials and external noise level analysis.

In Figure 3C, we can see the FAST filter's robustness to increasing levels of noise, with the edge weights picking up significant results at the correct time steps in almost all cases in the 100–250 trial range. At 50 trials, it still performs relatively well, but there is a decreasing performance at very high levels of external noise. The trend shows that as the number of trials increases, the detection ability improves while increasing noise decreases this detection ability. The mean edge weights in the unfiltered case fail to detect these simulated ERPs in all levels of noise and trial sizes.

The findings when we set the number of windows to the sampling rate exhibit promise, revealing that despite a reduction in performance when utilizing maximum temporal resolution with 200 windows in contrast to 10 temporal windows, as there is clear failure at higher levels of noise at 50 trials (although the ERP is detected at 50 trials at no external noise; $p$ = 0.04941), the decline is minimal particularly with an optimal number of trials. This observation underscores the robustness of FAST connectivity in variance to window length variations, highlighting its comparative advantage over existing methodologies in capturing DFC changes.

#### *Application to VSTM Binding in AD*

EEG microstates:
Transient patterns in the EEG that reflect brief periods of quasistable brain states, important for understanding basic neural processes and distinguishing between neurological conditions.

EEG microstates are transient patterns of the EEG that occur in very small temporal windows and are considered to be related to the most basic of human neurological processes. They have been previously shown to be able to distinguish between neurological disorders such as schizophrenia based on these tiny temporal window differences where the overall functional connectivity of the brain may be very similar (Koenig et al., 1999). Recently, there has been a significant interest in these EEG microstates in neurological disorder diagnosis.

We ran experiments for the MCI-FAM and MCI-SPO datasets separately with a single FAST filter computed from participants in both the shape and binding tasks for each frequency band. As mentioned in our Statistical Methods section, we computed the mean clustering coefficient and edge weights of each participant at 10 disjoint temporal windows of the total 1-s epoch of interest and undertook nonparametric statistical testing between controls and patients to look for temporal windows where there is concurrently a significant difference between controls and patients in the binding task and no significant difference in the shape task. This exploits the proposed binding deficit established in Pietto et al. (2016). Figure 4 shows the plots of the log of the $p$ values of the patients versus controls in shape and binding tasks for the MCI-FAM dataset, with values below the red line indicating a significant difference.

The first thing we notice is that the behavior of the delta and theta bands in the binding task is nearly identical, with significant results found at 0.3–0.4 s by the mean edge weights and 0.4–0.5 by the mean weighted clustering coefficient. We can see how having two different network metrics can aid the detection of clinically significant results. We see that the shape tasks for these time steps are not significantly different; thus, these can be considered results of clinical interest as the specificity of binding deficits observed behaviorally is replicated here at a neural level. The binding task in the alpha band seems to show some behavior similar to the theta and delta bands, with a clinically interesting result at 0.4–0.5 s; however, this would not pass FDR correction. The beta band seems to follow the same pattern in the 0.3–0.6 range, with a dip toward the significance line in the binding task and movement away from it in the shape task. In light of volume conduction effects, the gamma band was found to yield spurious results, consequently warranting its exclusion from our analysis. Overall, we notice a trend of clinically significant results in the 0.3- to 0.6-s range in the MCI-FAM dataset, mainly in the lower frequency bands.

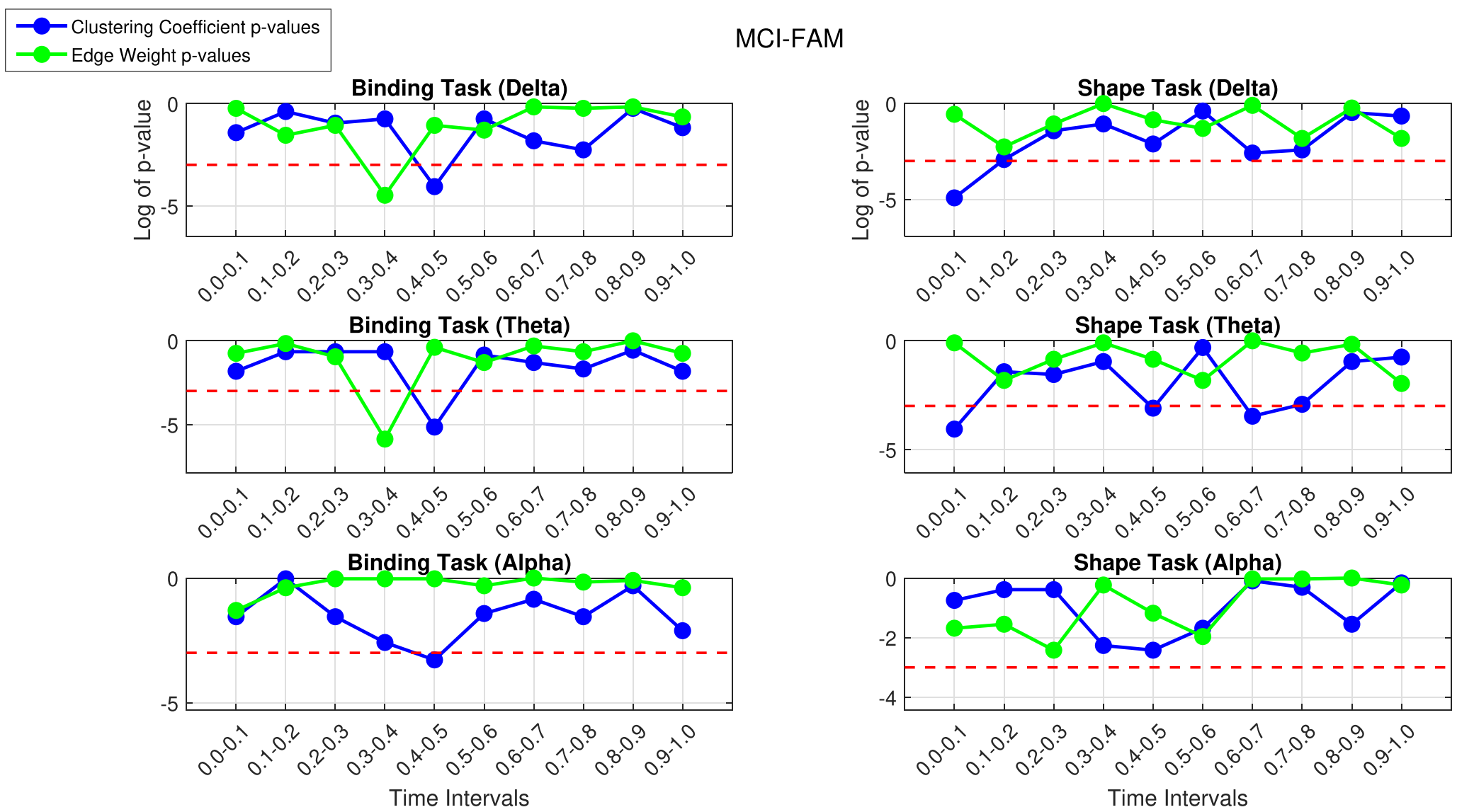

**Figure 4.** The *p* value plot for controls versus patients using FAST connectivity in the shape and binding task in the delta (top), theta (middle), and alpha (bottom) bands for MCI-FAM.

Figure 5 shows the plots for the log of the *p* values for patients versus controls in the shape and binding VSTM task for the MCI-SPO dataset against the time intervals.

We notice similar patterns in the lower frequency bands, with the binding task in the delta and theta bands following similar patterns in the mean edge weights and the network metrics picking up clinically interesting results in the theta band at 0.4–0.5 and 0.5–0.6 s. This is an overlap in the 0.3- to 0.6-s range with the MCI-FAM dataset. The delta band has a clinically interesting result at 0.5–0.6 s, with the mean edge weights and a highly significant result at 0.8–0.9 s using the weighted clustering coefficient; this could be related to the emotion-related late positive potential (LPP). The LPP, characterized by a gradual positive shift in activation, typically manifests approximately 400–1,000 ms following stimulus presentation. Its amplification has been linked to memory encoding and retention mechanisms (Pietto et al., 2016). Furthermore, it has been correlated with postretrieval phases, such as decisional and evaluation processes, which could be affected by AD. Again, the overlap with the delta band in the MCI-FAM group in the 0.3- to 0.6-s range is seen. Moreover, the alpha band has clinically interesting results at 0.1–0.2 s with the weighted clustering coefficient and 0.2–0.3 s and 0.5–0.6 s with the mean edge weights. It seems that the similarity to the theta band in the MCI-FAM group is growing with the behavior in the 0.3–0.6 s range being much more prominent in the alpha band. The beta band has a significant result at the 0.4- to 0.5-s time step, which is similar to the behavior of the beta band in the MCI-FAM dataset (dip toward significance line in binding task, movement away in the shape task). Overall, there are consistent overlapping results of clinical interest in the 0.3- to 0.6-s temporal range.

We then applied FDR correction at the 10% and 5% levels to account for multiple comparisons. Table 3 shows the time intervals at which there is a significant difference between controls and patients in the binding task and not in the shape task after applying FDR correction at the 5% and 10% levels.

The main results that survive FDR correction are 0.3–0.6 range results in the lower frequency bands. There are overlapping significant results in the 0.3- to 0.6-s range in the MCI-FAM and MCI-SPO datasets. The 0.3–0.6 results in the MCI-SPO alpha band-pass FDR correction, thus bringing evidence of an aging interplay between the alpha band frequency in the older MCI-SPO patient group mimicking the behavior of the theta band in the younger MCI-FAM patient group. The binding task effect sizes are all consistently greater in the patient group, suggesting increased FAST connectivity in Alzheimer's patients during binding VSTM tasks. This correlates to our time series plot of the FAST filtered mean patient and control matrices. The 0.3–0.4 range in the MCI-FAM theta band shows greater squared difference values in controls with a rapid switch to greater values in patients in the next time step. We conjecture that this could be due to a delay in the onset of the P300 in patients resulting in the increased FAST connectivity to only appear *after* the onset of the P300 in the controls in the previous time step.

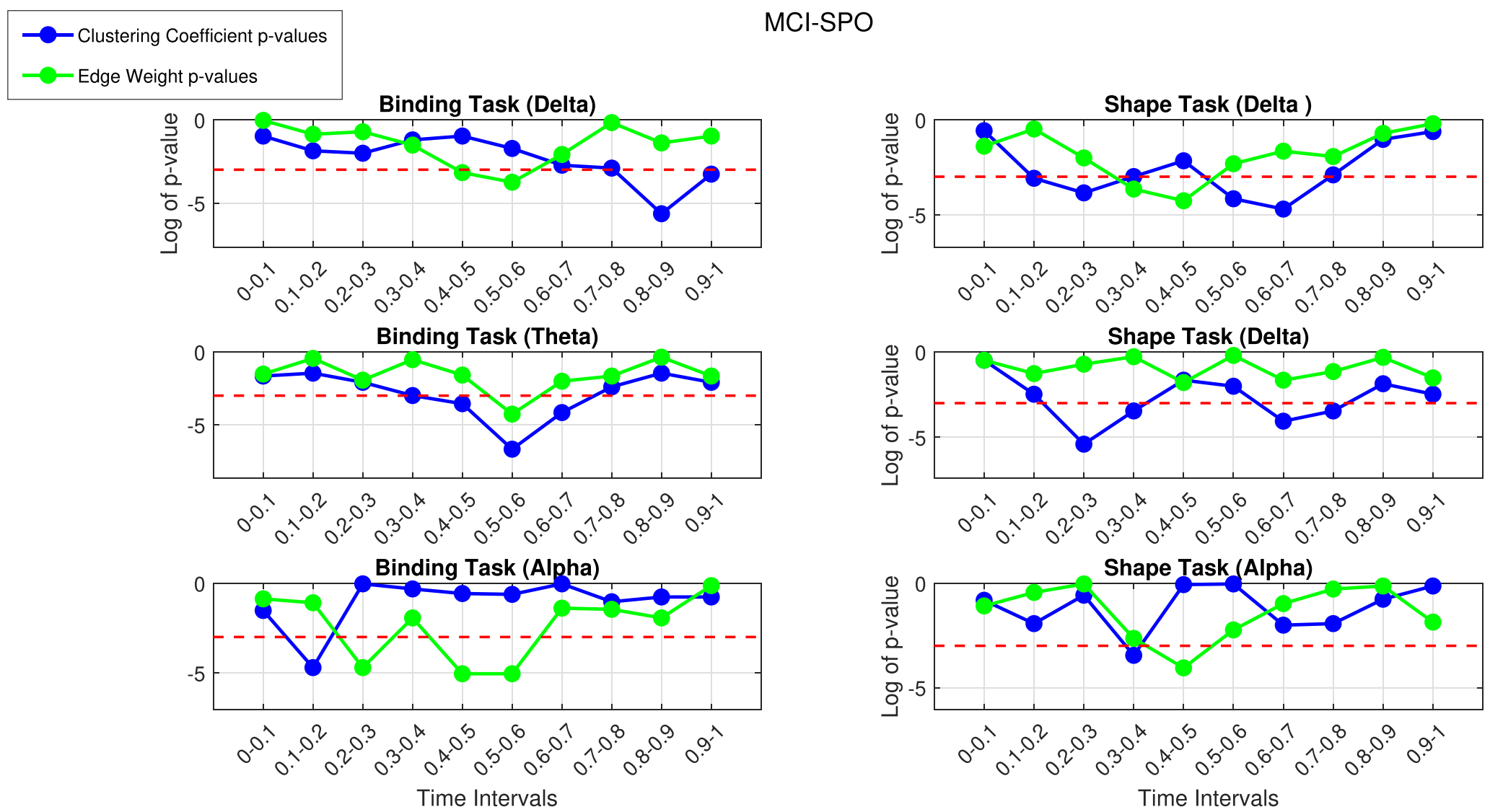

**Figure 5.** The *p* value plot for controls versus patients using FAST connectivity in the shape and binding task in the delta (top), theta (middle), and alpha (bottom) bands for MCI-SPO.

Furthermore, we computed the average of the FAST filtered connectivity tensors for all participants and controls in the binding task for both MCI-FAD and MCI-SPO. This gave us one tensor for patients and one for control, representing the general activity in terms of functional connectivity in the task.

We mapped the nodewise connectivity to specific electrodes and compared controls and patients during a specific time step in the P300 range; in particular, we set the time step corresponding to around the 0.5-s mark in the theta band in both datasets as this was the

**Table 3.** FDR corrected datasets (10% level)

| Freq. band | Range | Binding *p* value | Binding effect size |
|---|---|---|---|
| MCI-FAM theta | **0.3–0.4**, **0.4–0.5** | 0.028, 0.058 | 1.60, −1.12 |
| MCI-SPO delta | 0.8–0.9 | 0.035 | −0.55 |
| MCI-SPO theta | **0.5–0.6** | 0.013 | −1.19 |
| MCI-SPO alpha | 0.1–0.2, 0.2–0.3, **0.5–0.6** | 0.09, 0.03, 0.03 | −0.88, −1.17, −1.10 |

The underlined texts represent the binding *p* value below 0.05. The **bold** font indicates ranges in the first 300 ms of the P300.

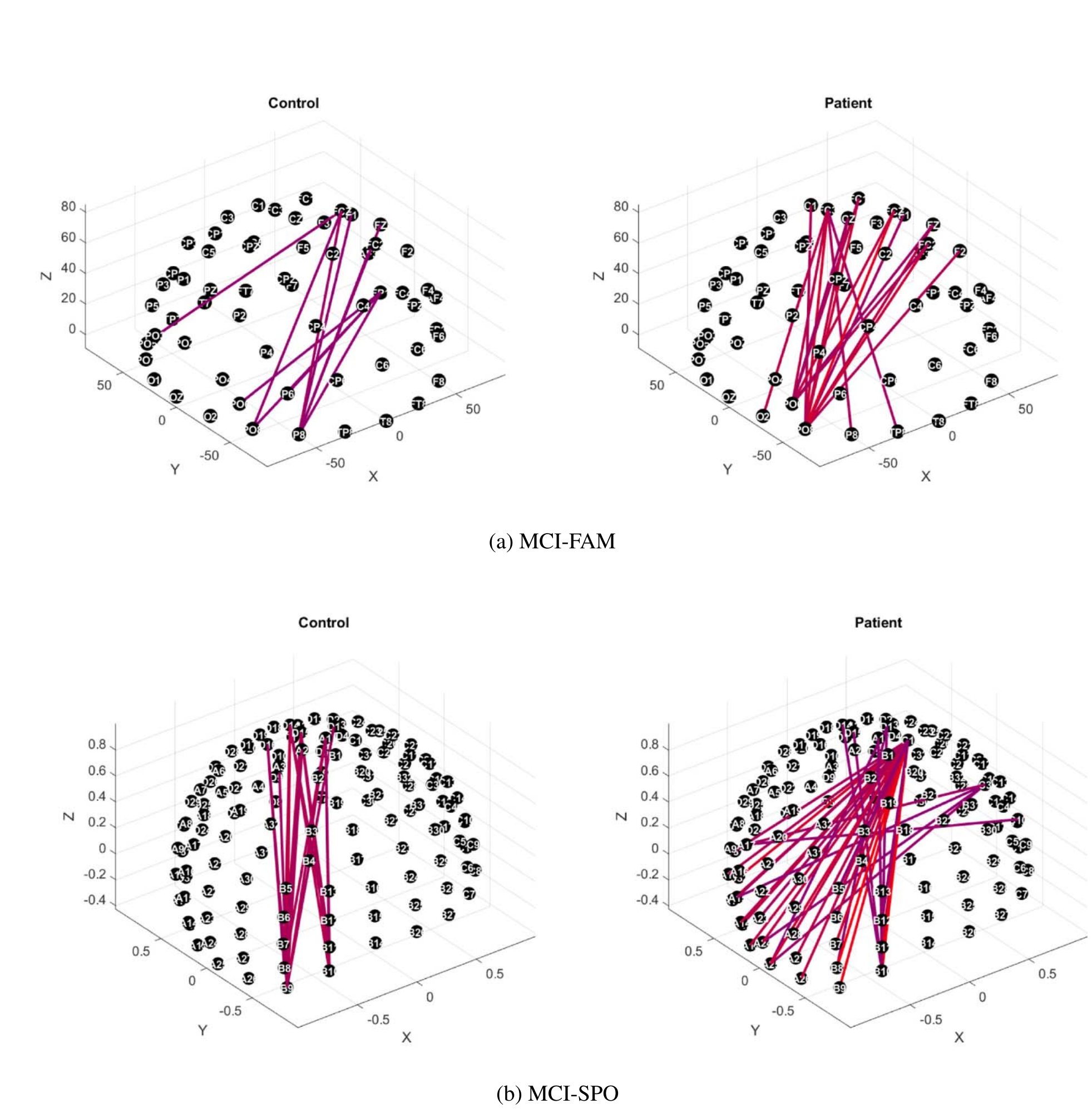

(a) MCI-FAM

(b) MCI-SPO

**Figure 6.** (A) Top 1% (MCI-FAM) and (B) top 0.1% (MCI-SPO) FAST connectivity topological scalp plots in the theta band (binding task) at 0.5 s poststimuli onset (intensity indicated on the color bar).

overlapping time step that was common to both datasets in the theta frequency band with similar effect sizes.

We noted that only the top 1% of connections in the union of both mean matrices for the binding task appeared in both controls and patients in MCI-FAD, while in MCI-SPO, there were similarly strong connections up to the top 0.1% of connections. This is due to both controls and patients in MCI-SPO having relatively large values for the instantaneous local variation with patients having more connections. In MCI-FAM, on the other hand, patients had typically greater local variation and more connections with a larger local Dirichlet energy compared with controls.

Figure 6 thus shows the connectivity plots where we compute the top 1% and 0.1% of strongest connections in the MCI-FAM and MCI-SPO mean matrices, respectively.

Interestingly, we notice an increase in instantaneous variability in patients in both MCI-FAD and MCI-SPO compared with controls in the P300 response. In particular, we can empirically observe the increased presence of strong potential "hubs" where multiple strongest connections start from the same node.

In MCI-FAD, these hubs originate from the PO3 and PO6 electrodes and are strongly connected to frontocentral regions. There is also a hub from FC3 to the parietal-occipital region. In general, we notice increased local variation from the parietal-occipital to the frontocentral region. In MCI-SPO, there is a clear major hub originating at C1 with large local variation, with regions across the entire parietal area; D1 is similar to C1 yet slightly less prominent, and smaller hubs also originate from across the parietal region. We notice that the difference between MCI-SPO and MCI-FAD being the increased local Dirichlet energy is more widespread across the parietal area.

These empirical results are fascinating but is not the focus of our study where we are more focused on global connectivity changes. The thorough broad spatial origins of these functional connectivity alterations, however, seem to be a promising avenue for future research.

## DISCUSSION

Our simulations showed the benefit of FAST connectivity compared with standard connectivity measures in picking up ERPs between participants with and without the discriminating ERP. We have provided a high-temporal resolution method that is robust to noise in small temporal windows while being very invariant to the window length. Thus, we achieve a better trade-off of noise robustness and temporal resolution compared with existing methods (Šverko et al., 2022).

After applying these results to our two independent MCI-SPO and MCI-FAM datasets, we found consistent overlapping clinically significant results in the 0.3- to 0.6-s range. This corresponds to the P300 range previously implicated in AD (Paitel et al., 2021; Parra et al., 2012). We also found evidence of altered functional connectivity related to increased localized signal variation in AD patients in the P300 range at time steps specific to the binding task. This supports the binding task deficit as a potential biomarker for AD. Theta and alpha band irregularities have been well researched to be linked to cognitive dysfunction and MCI due to AD (Babiloni et al., 2018; Klimesch, 1999; Schmidt et al., 2013). While the slowing of the alpha band is an indicator of progressing AD (Li et al., 2020). The alpha and theta band "shifts" could be of clinical significance as the two independent datasets differ by age; this frequency shift may be signaling age-related compensatory neural mechanisms, which have been previously reported during memory tasks performed in the fMRI scanner (Parra et al., 2013). More importantly, we note that our results correspond to what is observed at clinical performance level of the VSTM tasks, with binding task performance being significantly worse in patients and controls in both MCI-SPO and MCI-FAD. We have thus confirmed at a neuronal level what we see in practice.

It is of interest to note that the increased local Dirichlet energy from parietal to central–frontal regions is consistent in both types of AD compared with controls, this suggests functional connectivity changes could indicate AD progression before structural changes occur. These hubs could indicate neuronal-level abnormalities of excitation and inhibition that are shown to be associated with tau and amyloid beta in preclinical models of AD (Ranasinghe et al., 2022).

It is promising that our simulations used a lower electrode density (where we employed 31 electrodes) and FAST connectivity was still able to reliably pick up simulated ERPs at a relatively low number of trials. This brings further confidence to our AD data results that employ 64 and 128 electrodes for the MCI-FAM and MCI-SPO datasets, respectively. This indicates that the electrode density had a fairly negligible influence on our results. Recent research (Hatlestad-Hall et al., 2023) supports this suggesting performance on functional connectivity measures is negligible above 60 electrodes.

While FAST connectivity is not in any way meant to replace traditional power-based analysis of ERPs, it has significantly improved performance in this *specific* time-locked task simulation. It shows that it can be a reliable tool to analyze DFC changes related to ERPs at miniscule timescales.

The data we have utilized in this work are a small sample of two populations with different risk levels (MCI) for AD (E280A-PSEN1 Familial AD, with 100% risk; Acosta-Baena et al., 2011; Lopera et al., 1997) and sporadic MCI with an unknown risk. Previous work has shown that EEG connectivity can distinguish mutation carriers from controls with accuracy near 90% (Parra et al., 2017), a classification accuracy never reached via pure behavioral scores (Parra et al., 2010). It has also been shown that the EEG features linked to VSTM binding deficits of these patients across the two variants are indistinguishable (Parra et al., 2010). We have shown that FAST connectivity can distinguish the subtle time-frequency changes between MCI-SPO and MCI-FAM. Parra et al. (2024) also recently showed that a cost of binding (drop in performance on the shape–color binding condition relative to memory for shapes only) greater that 20% was associated to increased amyloid beta deposits in still cognitively unimpaired older adults.

In light of this, our application of FAST connectivity to the MCI-SPO and MCI-FAM datasets have provided results of interest for understanding deficits of VSTM binding in AD. With more data and further analysis (including at an individual, rather than just group level), this could potentially also be useful as a diagnostic indicator for the early detection of AD and the progression of MCI to dementia. Given the noninvasive nature of EEG signal analysis combined with the low computational cost of using GVD connectivity with a relatively small number of patients, we see the potential for this method to be used in the diagnosis of AD for low-income individuals. Furthermore, the task has been recommended by international consensus groups (Costa et al., 2017) as a promising preclinical test for AD. Some have already introduced the task in their clinical practice. The task has now been introduced in major international cohort studies such as PREVENT (Parra et al., 2021) and RedLAT (Ibanez et al., 2021).

Working memory tasks often require participants to engage in sustained cognitive effort, leading to potential cognitive fatigue, especially in prolonged experimental sessions. FAST connectivity shows potential to address this challenge by exploring the feasibility of achieving accurate ERP detection with a lower number of trials. Additionally, we should take into account economic considerations prevalent in low-income countries, where optimizing experimental protocols can significantly reduce costs associated with data acquisition and analysis.

An obvious application of FAST connectivity or similar methods would be in brain computer interfaces (BCIs; Daly et al., 2012). The ability to exploit discriminating information in real time from cheap, noninvasive EEG signals provides an avenue for a realistic, widely accessible BCI. While we are still far away from real-time detection, FAST connectivity, with its high invariance to window-length changes and performance at temporal resolution, shows potential for this one day being a possibility. Given recent advances in network-based BCIs (Gonzalez-Astudillo et al., 2021) and the proven importance of functional connectivity dynamics in the performance of BCIs (Daly et al., 2012), this would be a worthwhile avenue to explore.

Machine learning can be implemented on the network metrics of GVD connectivity due to the high-temporal resolution of the metrics; this could add important transient information to machine learning algorithms significantly improving performance akin to the wavelet transform shown to increase classification accuracy of neurological disorders by adding transient information (Kumar et al., 2022). Chu et al. (2023) also showed that combining graph metrics based on DFC in small temporal windows with typical classification algorithms showed significantly improve performance in the early detection of Parkinson's disease (PD), showing the discriminating ability of these EEG micro-states.

The diagnosis of neurological disorders in psychiatry is an area of uncertainty due to the overlap between disorders. FAST connectivity provides potential in the analysis of EEG signals to provide a more quantitative judgment on the nature of the neurological disorder.

While the results show the promise of this new methodology, it is worthwhile reflecting on where it may fail. We have already mentioned that it would not be suitable for picking up transient functional activity among connections that are otherwise independent and so having low long-term connectivity. Additionally, since the FAST filter is based on long-term connectivity, it should foremost be applied to singular cognitive processes. This means that it may not be suitable to apply to instances where there are expected changes in the cognitive function of a task, for example.

## CONCLUSION

We have introduced FAST connectivity, an algorithm that leverages a single global filter computed from both groups of participants in a given EEG paradigm. We have shown, in controlled simulations on synthetic data, that the method outperforms previous graph-variate and traditional power spectra methods in detecting subtle differences in small temporal windows between groups of participants with and without a simulated ERP in noisy conditions. We have also shown the lower dependence on window length of the method, providing an alternative to existing sliding window methods. Of notable interest is the fact that there is still relatively good performance when the window length is equal to the sampling rate, allowing us to potentially detect changes in temporal windows at a very granular timescale while also showing potential to reduce the required number of trials required for an ERP analysis. Applying FAST connectivity to two independent cohorts of sporadic and familial MCI patients engaged in VSTM tasks, we found significant differences between groups for time steps in the 0.3- to 0.6-s range in the binding task but not in the shape task; previous studies correspond this to the P300 ERP range. This was more prominent in the lower frequency bands, in particular, the theta band, corresponding with previous studies on the binding deficit and the role of the theta band in AD and general dysfunction in memory and cognition. Future work should aim to focus on further studying the spectral properties of the FAST filter, such as the spectral profile of the graph Laplacian in order to analytically understand its noise reduction and important connection promotion effects. Different instantaneous node functions based on the structure of the graph signal data should also be explored.

## ACKNOWLEDGMENTS

I.A. is partially supported by grants from ANID/FONDECYT Regular (1210195 and 1210176 and 1220995), ANID/FONDAP/15150012, ANID/PIA/ANILLOS ACT210096, ANID/FONDAP 15150012, and the MULTI-PARTNER CONSORTIUM TO EXPAND DEMENTIA RESEARCH IN LATIN AMERICA (ReDLat, supported by Fogarty International Center [FIC] and National Institutes of Health, National Institutes of Aging [R01 AG057234, R01

AG075775, R01 AG21051, CARDS-NIH], Alzheimer's Association [SG-20-725707], Rainwater Charitable foundation – Tau Consortium, the Bluefield Project to Cure Frontotemporal Dementia, and Global Brain Health Institute). The contents of this publication are solely the authors' responsibility and do not represent the official views of these institutions. The funders had no role in study design, data collection and analysis, publication decisions, or manuscript preparation.

## AUTHOR CONTRIBUTIONS

Om Roy: Conceptualization; Formal analysis; Investigation; Methodology; Project administration; Validation; Visualization; Writing – original draft; Writing – review & editing. Yashar Moshfeghi: Supervision; Writing – review & editing. Agustin Ibanez: Data curation; Resources. Francisco Lopera: Data curation; Resources. Mario Parra: Data curation; Validation; Writing – review & editing. Keith Smith: Supervision; Validation; Writing – review & editing.

## FUNDING INFORMATION

Om Roy is supported by the Engineering and Physical Sciences Research Council (EPSRC) Student Excellence Award (SEA) Studentship provided by the United Kingdom Research and Innovation (UKRI) council, RTSG Grant Number: 12212S220211-116.

## DATA AVAILABILITY STATEMENT

The data that supports the findings of this study are available from the corresponding author upon reasonable request.